\documentclass[aps,twocolumn,prb,showpacs]{revtex4}
\usepackage{graphics,amssymb,amsmath,epsfig}
\begin{document}

\title{Autonomous Stochastic Resonance in Fully Frustrated Josephson-Junction Ladders}
\author{Gun Sang Jeon}
\affiliation{Center for Strongly Correlated Materials Research,
Seoul National University, Seoul 151-747, Korea}
\author{M.Y. Choi}
\affiliation{Department of Physics, Seoul National University, Seoul 151-747,
Korea}

\begin{abstract}
We investigate autonomous stochastic resonance in fully frustrated
Josephson-junction ladders, which are driven by uniform constant currents.
At zero temperature large currents induce oscillations between the
two ground states, while for small currents the lattice potential forces
the system to remain in one of the two states.
At finite temperatures, on the other hand, oscillations between
the two states develop even below the critical current;
the signal-to-noise ratio
is found to display array-enhanced stochastic resonance.
It is suggested that such behavior may be observed experimentally
through the measurement of the staggered voltage.
\end{abstract}


\maketitle

The stochastic resonance (SR), which describes the phenomenon
that a noise of finite strength optimizes the response signal to a weak
periodic external force in a nonlinear system, has attracted much
attention for the past two decades.\cite{review}
Conventional SR phenomena are generally known to require an energy barrier
between two stable states, a weak external periodic force, and noise.
However, the SR behavior was also exhibited without any external periodic force,
in limit-cycle systems\cite{Gang} and excitable systems.\cite{CR,CR2}
Subsequently such coherence resonance or {\em autonomous} SR
has also been observed in systems with
delay\cite{delay} and inertia.\cite{inertia}
In the latter cases, SR phenomena can still be understood by the
time-scale matching argument since the delay or the inertia
provides an external time-scale.
In the former, on the other hand, such an external time scale
apparently does not exist,
and different noise dependencies of the activation
time and the excursion time have been proposed as
an explanation of the coherence resonance.\cite{CR2}
Recently, the fully-frustrated Josephson-junction ladder has been
proposed to give a good physical realization of the standard two-state SR system
with many degrees of freedom.\cite{acSR}
In the presence of the external current which is periodic in time and staggered
in space, the SR behavior and other rich physics
have been demonstrated theoretically in the ladder system.
Furthermore, recent progress in the fabrication technique makes such
Josephson-junction systems available for experimental study.\cite{review2,Delsing}

In this work we study fully frustrated Josephson-junction
ladders driven by {\em uniform constant} currents,
paying particular attention to the possibility of the autonomous SR phenomena.
Note that this system is {\em bistable}, possessing two stable states;
this is in sharp contrast with the usual excitable system,
characterized by a single stable state.
Numerical integration of the coupled equations of motion,
established within the resistively shunted junction (RSJ) model,
shows the existence of the critical current beyond which
oscillations emerge between the two ground states. Below the
critical current, such oscillations are suppressed by the lattice
potential; here the addition of noise currents, relevant at finite
temperatures, induces again oscillations, giving rise to SR
behavior. In particular, the signal-to-noise ratio (SNR) is found
to display array-enhanced SR phenomena, which may be observed
experimentally through the measurement of the staggered voltage.
To our knowledge, such array-enhanced SR behavior has not been
observed in bistable system without periodic driving although
coherence resonance was reported in the bistable regime of the
FitzHugh-Nagumo model.~\cite{bistable}

\begin{figure}
\epsfig{file=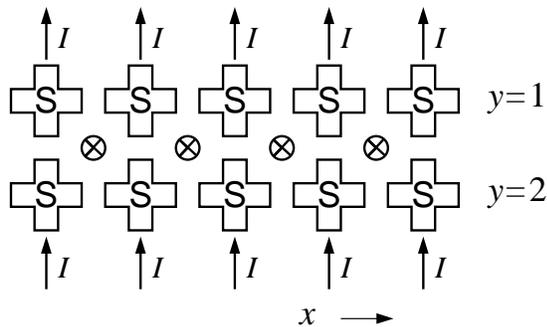,width=7.5cm}
\caption{Schematic diagram of a fully frustrated Josephson junction ladder
driven by uniform constant currents.}
\label{fig:ladder}
\end{figure}

The RSJ model, with single-junction
critical current $I_c$ and shunt resistance $R$,
is described by the set of equations of motion
for the phase $\phi_i$ of the superconducting order parameter
on grain $i$:\cite{acSR,review2}
\begin{equation}
\label{eqmotion}
{\sum_j}' \left[ {\hbar \over 2eR} {d \phi_{ij} \over dt}
+ I_c \sin(\phi_{ij}-A_{ij}) + \eta_{ij}\right]
= I_i^{\rm ext} ,
\end{equation}
where $\phi_{ij}\equiv \phi_i - \phi_j$ is the phase difference
across the junction $(ij)$,
the thermal noise current $\eta_{ij}$ is assumed to be the white noise
satisfying
\begin{equation}
\langle \eta_{ij}(t{+}\tau) \eta_{kl} (t) \rangle
= {2 T \over R} \delta (\tau) (\delta_{ik}\delta_{jl} -
\delta_{il}\delta_{jk}),
\end{equation}
and the primed summation runs over the nearest neighbors of grain $i$.
We choose the Landau gauge and write the gauge field $A_{ij}$,
which describes the transverse magnetic field, in the form
\begin{equation}
A_{ij} = \left\{
\begin{array}{ll}
0 &  \hbox{for }(ij) \hbox{ legs},\\
2\pi f x &  \hbox{for }(ij) \hbox{ rungs},
\end{array}
\right.
\end{equation}
where $f \equiv \Phi / \Phi_0$ is the flux $\Phi$ per plaquette
in units of the flux quantum $\Phi_0 \equiv hc / 2e$
and $x \, (= 1,2, \ldots, L)$ denotes the position along the leg.
As illustrated in Fig.~\ref{fig:ladder}, the ladder, placed along the $x$ direction,
is driven by the uniform constant currents $I$ along the rung direction (i.e.,
the $y$ direction):
The external current $I_i^{\rm ext}$ fed into grain $i$ is thus given by
\begin{equation}
I_i^{\rm ext}(t) = (-1)^y I ,
\end{equation}
where $y \, (= 1,2)$ is the leg index of the $i$th grain.
Henceforth, we write the current, the temperature, and the time in units of
$I_c$, $\hbar I_c / 2e$, and $\hbar / 2eRI_c$, respectively.

\begin{figure}
\epsfig{file=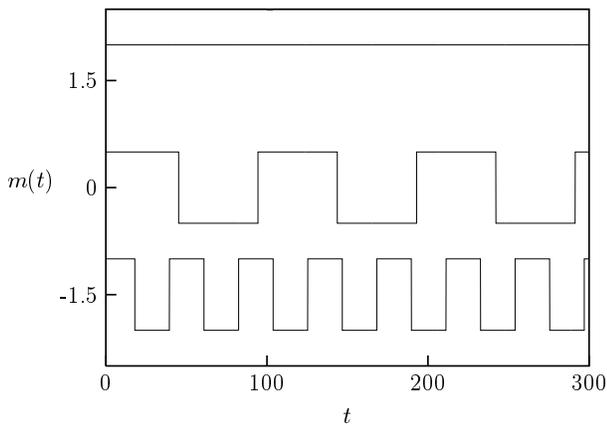,width=8cm}
\vspace*{0.4cm}
\caption{Time evolution of the staggered magnetization for various
currents $I = 0.2,\, 0.25$, and $0.3$ from above.
For clarity, the data corresponding to $I=0.2$ and $0.3$ have been shifted by $1.5$
upward and downward, respectively.}
\label{fig:zeroT}
\end{figure}
In the absence of external currents, the fully frustrated ($f=1/2$)
ladder is well known to have two degenerate ground states,
where a vortex is located on every two plaquettes.\cite{acSR}
(Note that the length $L$ of a fully frustrated ladder should be an even number.)
These states can be characterized by the {\em staggered magnetization}
\begin{equation}
m \equiv {1 \over L} \sum_X (-1)^X \left(n_X - {1\over2} \right) ,
\end{equation}
where $n_X$ is the vorticity on plaquette $X$. The vorticity is
given by the plaquette sum $n_X = (1/2\pi) \sum_{P_X}
(\phi_i{-}\phi_j{-}A_{ij} )+ 1/2$, where the gauge-invariant phase
difference is defined modulo $2\pi$ in the range $(-\pi,\pi]$.
Here uniformly applied currents in the $y$ direction drive
vortices to the $x$ direction, generating a flow of the vortex
array along the ladder. Due to the spatial periodicity of the
vortex array, the flow in the $x$ direction results in the
alternation between the two ground states, which can be manifested
by the oscillation of the staggered magnetization. In contrast to
a homogeneous thin film, the array system has a lattice potential
which tends to suppress the flow of a vortex array. Accordingly,
oscillations of the staggered magnetization should appear only
when the external currents exceed the critical value $I_0
=\sqrt{5}{-}2$.\cite{I0}
The time evolution of the staggered magnetization at zero temperature,
obtained from direct integration of Eq.~(\ref{eqmotion})
and presented in Fig.~\ref{fig:zeroT} for various values of $I$,
indeed confirms the existence of the critical value:
For small currents ($I=0.2 < I_0$), the staggered magnetization remains constant,
indicating that the ladder is pinned by the lattice potential
to one of the two ground states.
Above the critical value $I_0$,
there arise oscillations of the staggered magnetization.
Note also that the oscillation frequency increases with currents,
reflecting that the vortex array flows faster at larger currents.

To probe the possibility of autonomous SR here, we consider the
average rate of the change in the staggered magnetization
\begin{equation}
v(t) \equiv {\delta m \over \delta t} \equiv {m(t) - m(t{-}\delta
t) \over \delta t} ,
\end{equation}
which measures the transition rate between the ground states. The
presence of periodic oscillations between the ground states is
manifested by the peak in the power spectrum $S_v(\omega)$ of
$v(t)$ at the corresponding frequency. We then compute the SNR:
\begin{equation}
R \equiv 10 \log_{10} \left[ {S \over N} \right],
\end{equation}
where the signal $S \equiv S_v(\omega{=}\omega_p )$ is the peak
value of the power spectrum (at the peak frequency $\omega_p$).
The background noise level $N$ is estimated by the power spectrum
averaged over the interval of frequency much higher than
$\omega_p$, where the power spectrum is flat.
Equation~(\ref{eqmotion}) has thus been integrated directly with
the time steps $\Delta t = 0.05$, from which $v(t)$ has been
computed. We have also reduced $\Delta t$ to $0.01$, to obtain the
same results within the numerical accuracy, and followed an
annealing schedule with the equilibration time $1000$ at each
temperature. The data have been averaged over $200$ to $5000$
independent runs, depending on the system size. For convenience,
$\delta t$ has been set to be $0.25$.
We have also considered smaller values of $\delta t$, and found
that the overall results remain qualitatively the same although
higher noise levels make it more formidable to perform precise
numerical analysis.  The periodic boundary conditions have been
imposed along the $x$ direction.
\begin{figure}
\epsfig{file=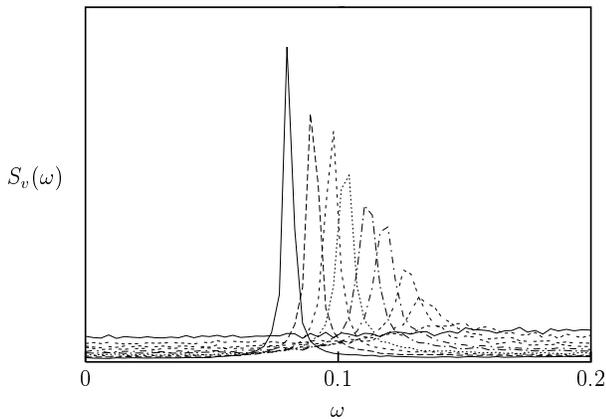,width=8cm} \vspace*{0.4cm} \caption{Power
spectrum of the transition rate $v(t)$ (in arbitrary units) in a
ladder of length $L=50$ for $I=0.25$ at various temperatures $T=
0.01,\, 0.02,\, 0.03,\, 0.04$, $0.06,\, 0.08,\, 0.12,\, 0.16,\,
0.24$, and $0.4$ from left. } \label{fig:PSabove}
\end{figure}
\begin{figure}
\epsfig{file=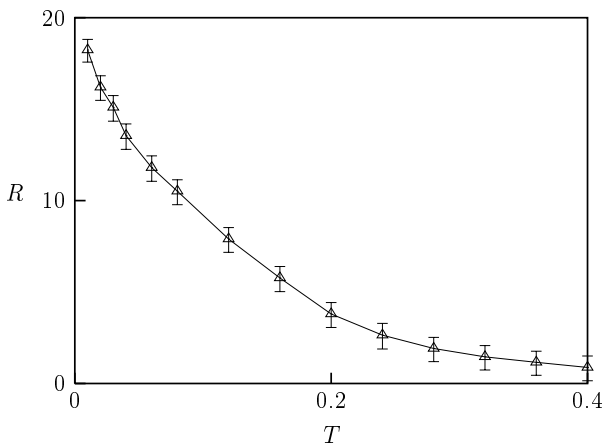,width=8cm} \caption{SNR versus temperature
in a ladder of length $L=50$ for $I=0.25$. Here and in other
figures applicable, error bars have been estimated by the standard
deviation, and the lines connecting data points are merely guides
to eyes. } \label{fig:SNRabove}
\end{figure}

We first examine the effects of noise above the critical current.
Figure~\ref{fig:PSabove} shows the power spectrum at
various temperatures in the ladder of length $L=50$
above the critical current ($I=0.25$).
The power spectrum exhibits a sharp peak even at zero temperature,
arising from the current-induced oscillation.
As the temperature is raised, however, the signal $S$ decreases monotonically,
reflecting that the thermal noise disturbs the
coherent motion of the vortex array.
Accordingly, the SNR displays monotonic decrease
with the temperature, as shown in Fig.~\ref{fig:SNRabove}.
\begin{figure}
\epsfig{file=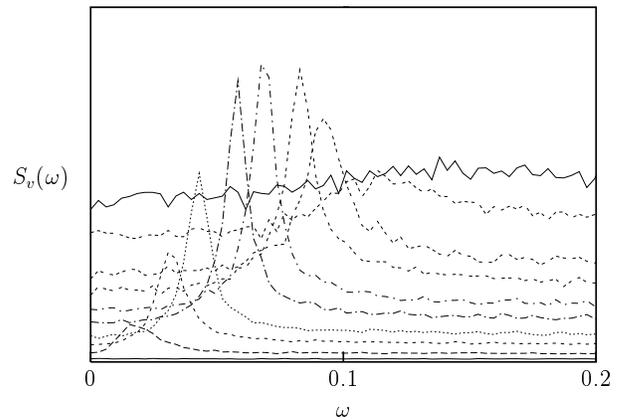,width=8cm} \vspace*{0.4cm} \caption{Power
spectrum of $v(t)$ (in arbitrary units)
in a ladder of length $L=50$ for $I=0.2$ at various temperatures
$T= 0.01,\, 0.02,\, 0.03,\, 0.04,\, 0.06,\, 0.08,\, 0.12,\, 0.16,\,
0.28$, and $0.4$ from below.
}
\label{fig:PSbelow}
\end{figure}

\begin{figure}
\epsfig{file=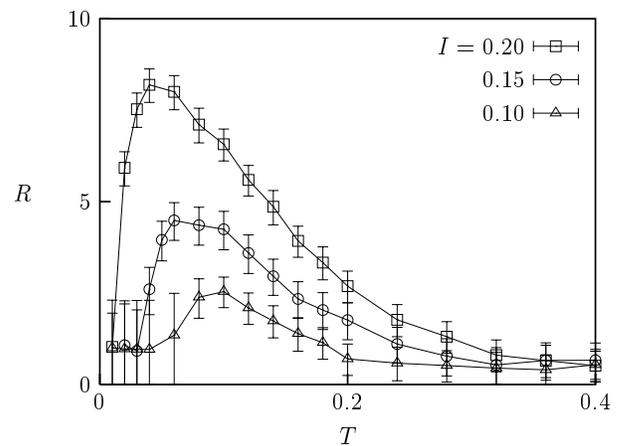,width=8cm}
\vspace*{0.4cm}
\caption{SNR versus temperature in a ladder of
length $L=50$ for various currents below $I_0$.}
\label{fig:SNR-I}
\end{figure}

\begin{figure}
\epsfig{file=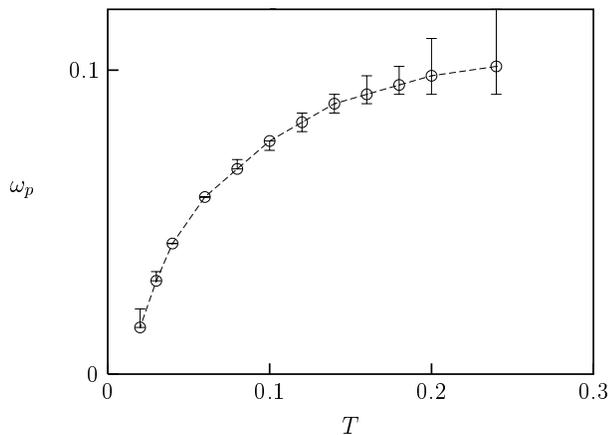,width=8cm}
\vspace*{0.4cm}
\caption{Peak frequency $\omega_p$
as a function of the temperature in a ladder of length $L=50$ for $I=0.2$.
}
\label{fig:peakfreq}
\end{figure}

We next reduce the current $I$ to $0.2$, below $I_0$. Unlike at
zero temperature, there emerge oscillations between the two ground
states at finite temperatures, as manifested by the peak of
$S_v(\omega)$ in Fig.~\ref{fig:PSbelow}. The signal $S$ increases
as the temperature is raised from zero while further increase of
the temperature tends to suppress the peak, eventually forcing the
power spectrum to be white. The SNR plotted in
Fig.~\ref{fig:SNR-I} for $I=0.2$ and lower values clearly displays
autonomous SR, i.e., SR without periodic
driving.\cite{Gang,CR,CR2}
Here it should be stressed that the SR observed above does not
describe the direct response to the {\em external} dc current.
As shown in Fig.~\ref{fig:peakfreq}, the peak frequency, which
is proportional to the mean flow velocity of the vortex array,
increases monotonically with the temperature, exhibiting no SR behavior.
This is consistent with the previous study
which concluded that the mobility of an overdamped particle
in a washboard potential does not display SR behavior.\cite{washboard}

\begin{figure}
\epsfig{file=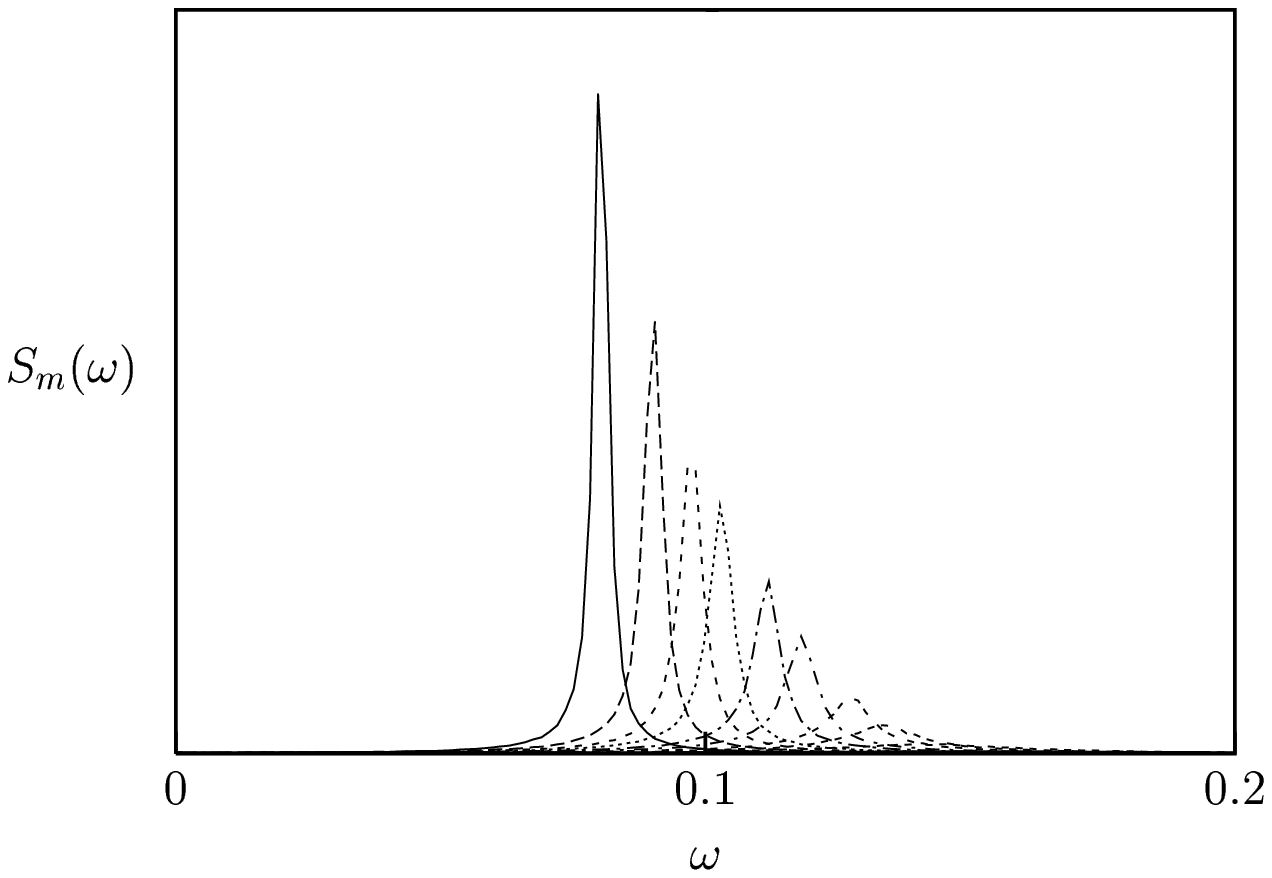,width=8cm} \centerline{(a)}
\epsfig{file=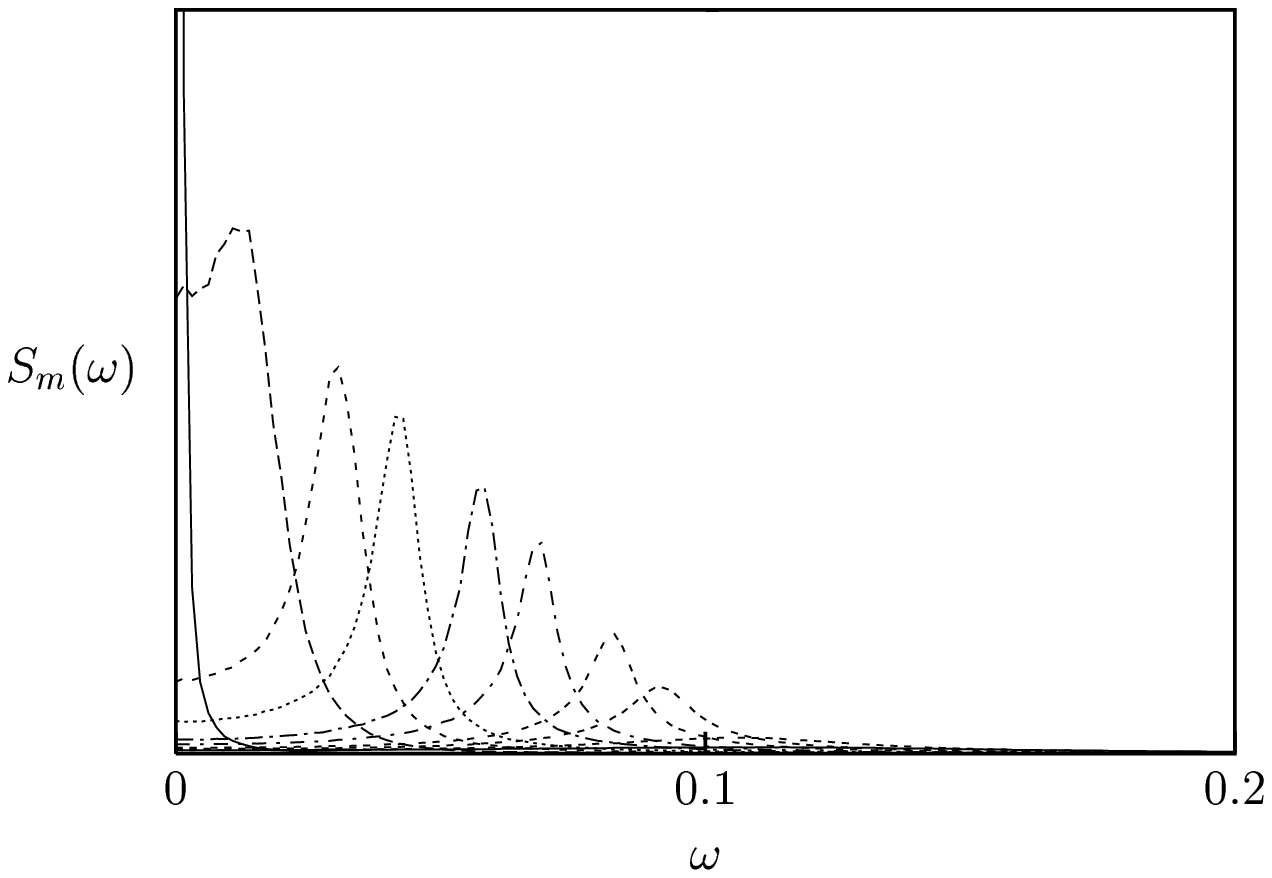,width=8cm} \centerline{(b)} \caption{ Power
spectrum of the staggered magnetization $m(t)$ (in arbitrary
units) at various temperatures $T= 0.01,\, 0.02,\, 0.03,\, 0.04$,
$0.06,\, 0.08,\, 0.12,\, 0.16,\, 0.24$, and $0.4$ (from left) in a
ladder of length $L=50$ for (a) $I=0.25$ and (b) $I=0.2$. }
\label{fig:PSm}
\end{figure}

Similar SR phenomena are also observed in the behavior of the
staggered magnetization $m(t)$ itself.  Figure \ref{fig:PSm}
displays the power spectrum $S_m(\omega)$ of the staggered
magnetization $m(t)$ both above and below the critical current.
Above the critical current, indeed similarly to $S_v(\omega)$
shown in Fig.~\ref{fig:PSabove}, $S_m(\omega)$ exhibits a sharp
zero-temperature peak at a finite frequency, which gets broadened
gradually with the introduction of thermal noise. Below the
critical current, on the other hand, the zero-temperature peak
develops at zero frequency ($\omega=0$), indicating the absence of
oscillation in the staggered magnetization.  As the temperature is
raised from zero, the dc component $S_m(\omega{=}0)$ becomes
smaller; instead there emerges a peak at a finite frequency
$\bar{\omega}_p$, which, like the peak frequency $\omega_p$ of
$S_v (\omega)$ (see Fig.~\ref{fig:peakfreq}), depends on the
temperature.  These behaviors of $S_m (\omega)$ and those of $S_v
(\omega)$ are in general consistent with the relation $S_v
(\omega) \propto \omega^2 S_m (\omega)$, accurate for small
$\delta t$.  Note here that such development of a finite-frequency
peak at finite temperatures is an indication of the coherence
resonance phenomena observed in the bistable regime of the
Fitzhugh-Nagumo model.~\cite{bistable}

\begin{figure}
\epsfig{file=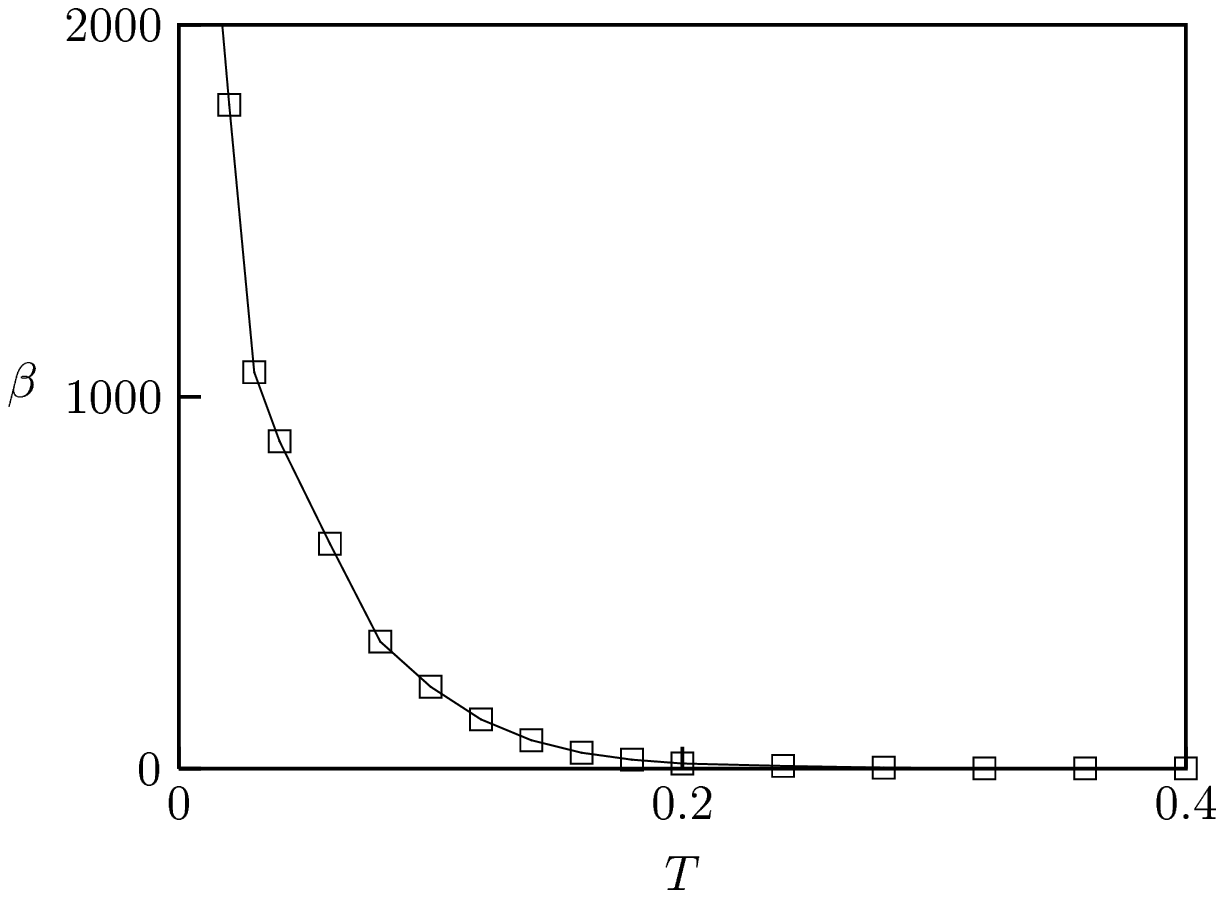,width=8cm} \centerline{(a)}
\epsfig{file=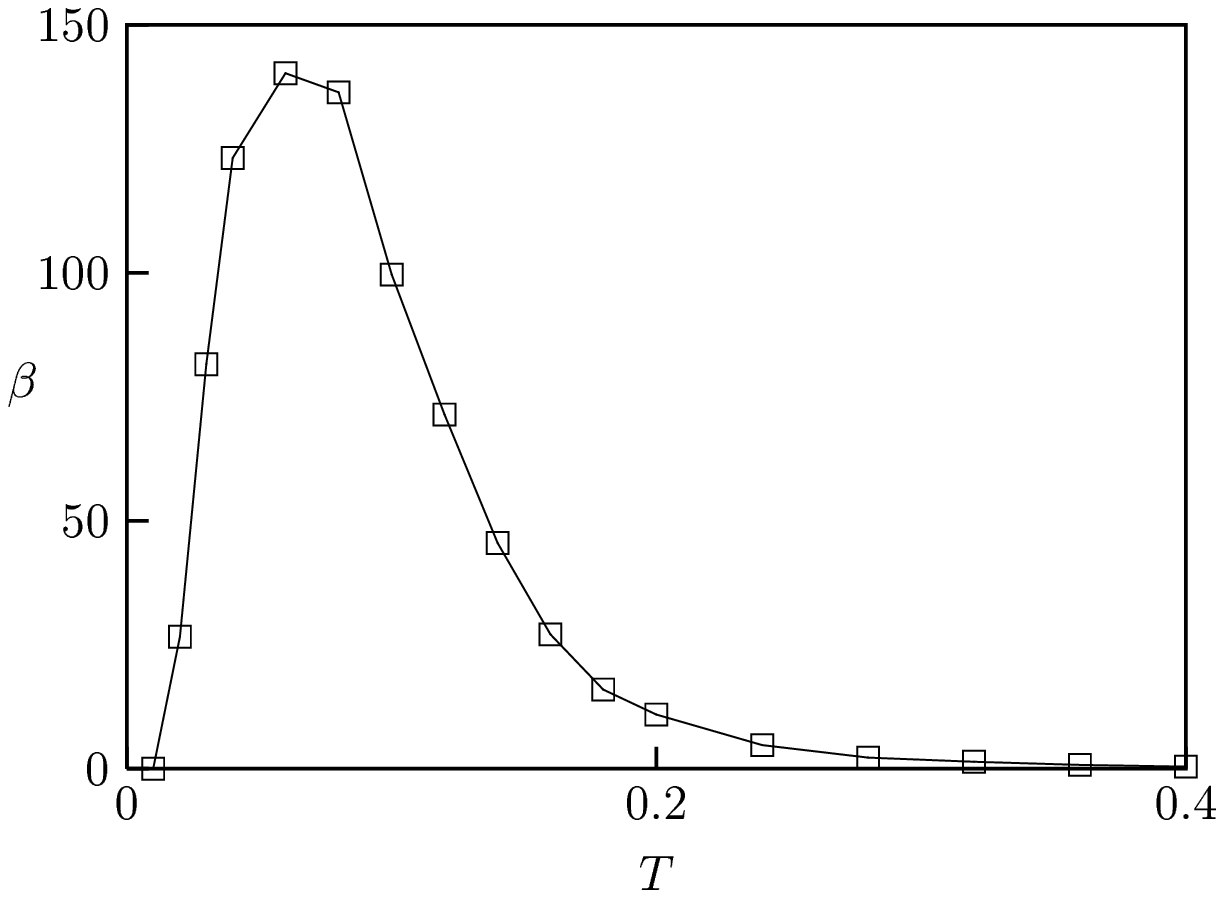,width=8cm} \centerline{(b)}
\caption{Coherence measure versus temperature in a ladder of
length $L=50$ for (a) $I=0.25$ and (b) $I=0.2$. } \label{fig:beta}
\end{figure}

To quantify conveniently the expected SR behavior of $m(t)$, we
compute the coherence measure\cite{Gang,beta}
\begin{equation}
\beta \equiv S_m(\bar{\omega}_p) \left({\Delta\omega \over
\bar{\omega}_p} \right)^{-1},
\end{equation}
where the half-width of the peak is given by $\Delta\omega \equiv
\omega_r - \bar{\omega}_p$ with $\omega_r \,(> \bar{\omega}_p)$
satisfying $2 S_m(\omega_r) = S_m(\bar{\omega}_p)$.  The coherence
measure $\beta$, which is plotted as a function of temperature in
Fig.~\ref{fig:beta}, also reveals the nature of the autonomous SR
behavior in the system.  Above the critical current it decreases
exponentially with the temperature [see Fig.~\ref{fig:beta}(a) for
$I=0.25$] while below the critical current a prominent peak
appears at a finite temperature, as demonstrated in
Fig.~\ref{fig:beta}(b) for $I=0.2$.  Thus confirmed is the
autonomous SR behavior of the staggered magnetization $m(t)$.

The autonomous SR in the fully frustrated ladder is rather peculiar,
arising from the combination of the external uniform currents and
the spatial periodicity of the ground state inherent in the system.
%
Although currents smaller than $I_0$ do not induce the vortex
array to move, they add an overall gradient to the lattice
potential along the $x$ direction. At finite temperatures, the
additional noise currents assist vortices to hop to neighboring
plaquettes while the potential gradient generated by the external
currents makes hopping in one direction more favorable than that
in the other. Consequently, the vortex array is encouraged to flow
to one direction, leading to oscillations of the transition rate
$v(t)$.  These oscillations grow with the temperature, which gives
the enhancement of the signal. At higher temperatures, however,
the lattice potential gradient can be neglected, and the random
nature of the noise disturbs the coherent flow, weakening the
signal. In this manner the two different roles of the noise
produces the autonomous SR in the system,\cite{comm} which also
provides a natural explanation of the current-dependence of the
SNR behavior in Fig.~\ref{fig:SNR-I}: At small currents the
lattice potential barrier between neighboring sites remain rather
high, and the noise required to overcome the barrier should be
strong. Accordingly, as the applied currents are reduced, the SR
temperature is expected to increase while the SNR peak diminishes
due to the large incoherent contributions of the noise.
\begin{figure}
\epsfig{file=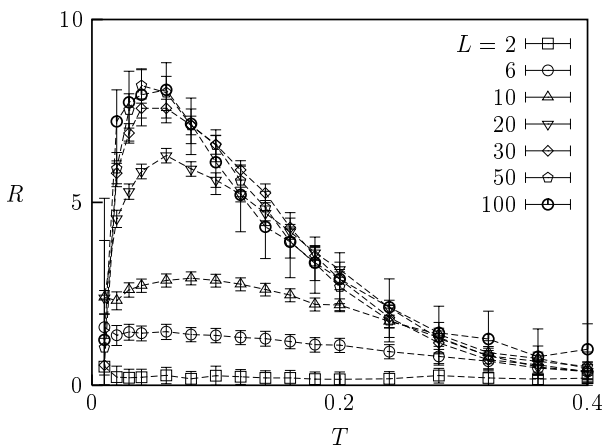,width=8cm}
\vspace*{0.4cm}
\caption{SNR versus temperature for various sizes and $I=0.2$.}
\label{fig:SNR-L}
\end{figure}

\begin{figure}
\epsfig{file=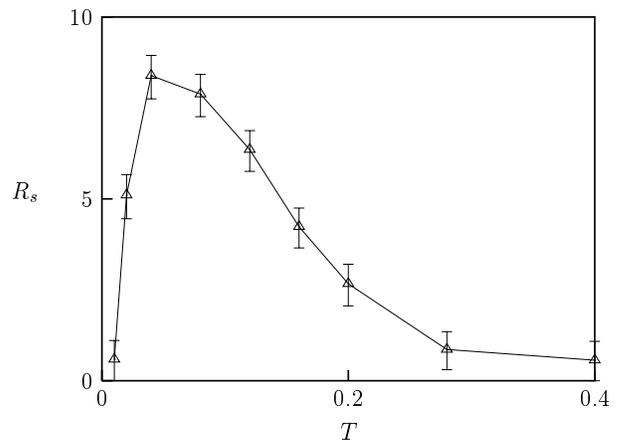,width=8cm}
\caption{SNR for the staggered voltage as a function of the temperature
in a ladder of length $L=50$ for $I=0.2$.}
\label{fig:SNR-Vs}
\end{figure}
Figure~\ref{fig:SNR-L}, showing the SNR versus temperature for various lengths,
discloses another interesting feature of the SR behavior in the ladder system.
In the ladder of length $L=2$, which is the minimum length required
for a fully frustrated ladder, the SNR exhibits no sign of SR.
As the length is increased, there first appears a broad peak in the SNR, which
then develops into a prominent peak.
Such {\em array-enhanced} SR phenomena
reflect the stronger coherence between vortices in longer ladders.
However, the enhancement should not persist with the length
since the ladder does not evolve long-range order in the thermodynamic limit.
Indeed the SNR in Fig.~\ref{fig:SNR-L} eventually saturates for $L \gtrsim 30$,
appearing independent of the length.
While array-enhanced resonance was first reported in the array
systems exhibiting conventional SR,\cite{array}
recent studies have also revealed the possibility of array-enhanced
coherence resonance in coupled excitable systems.\cite{array2}
It is remarkable that similar array-enhancement of autonomous SR can also be
observed in the Josephson-junction ladder, which possesses two stable states
like the conventional SR system (but in the absence of periodic driving).
%

Finally, we discuss how to observe the autonomous SR in experiment.
As a candidate for the measurable quantity characterizing the
autonomous SR, we suggest the average staggered voltage:
\begin{equation}
V_s (t) \equiv {1 \over L} \sum_x (-1)^x \bar{V}_x(t),
\end{equation}
where $\bar{V}_x(t)$ is the rung voltage at position $x$ averaged
over the interval $\delta t$ around time $t$. The rigid motion of
the vortex array at low temperatures should give rise to periodic
oscillations of $V_s(t)$, similarly to the transition rate $v(t)$.
The SNR for the staggered voltage $V_s(t)$, denoted by $R_s$, is
plotted in Fig.~\ref{fig:SNR-Vs}, where SR behavior similar to
that of $v(t)$ is shown. We thus suggest that the autonomous SR
phenomena in the ladder should be observed through the measurement
of the staggered voltage.

In summary we have investigated fully frustrated Josephson-junction
ladders, driven by uniform constant currents,
with regard to the possibility of stochastic resonance.
While the system under large currents displays oscillations between the
two ground states, the lattice potential suppresses
such oscillations for small currents.
Still the addition of noise currents, relevant at finite temperatures,
induces again oscillations, giving rise to stochastic resonance behavior.
In particular the signal-to-noise ratio has been found to display
array-enhanced stochastic resonance phenomena.
It has also been suggested that such behavior may be observed experimentally
through the measurement of the staggered voltage.
In view of this, the fully frustrated Josephson-junction ladder
makes a good physical realization of the system with many degrees of freedom,
adequate for the study of the autonomous
as well as the conventional stochastic resonance phenomena.

We thank P.\ Minnhagen and B.J.\ Kim for useful discussions.
This work was supported by the Korea Science and Engineering
Foundation through the Center for Strongly Correlated Materials
Research (GSJ) and by the Ministry of Education through the BK21 Project (MYC).

%


\begin{thebibliography}{99}
\bibitem{comm} Similar argument was also given as an
explanation of the autonomous SR in the limit-cycle system (see
Ref.~\onlinecite{Gang}).
\end{thebibliography}
\end{document}